\documentstyle[preprint,aps]{revtex}

\begin{document}
\draft
\title{Preparation of pseudo-pure states by line-selective pulses in Nuclear
Magnetic Resonance \thanks{%
e-mail: trap@nmr.whcnc.ac.cn; Fax: 0086-27-87885291.}}
\author{Xinhua Peng$^1$, Xiwen Zhu$^1$, Ximing Fang$^{1,2}$, Mang Feng$^1$, Kelin Gao%
$^1$, and Maili Liu$^1$}
\address{$^1$Laboratory of Magnetic Resonance and Molecular\\
Physics, Wuhan Institute of Physics and Mathematics, the Chinese Academy of\\
Sciences, Wuhan, 430071, People's Republic of China \\
$^2$Department of Physics,\\
Hunan Normal University, Changsha, 410081, China}
\maketitle

\begin{abstract}
A new method of preparing the pseudo-pure state of a spin system for quantum
computation in liquid nuclear magnetic resonance (NMR) was put forward and
demonstrated experimentally. Applying appropriately connected line-selective
pulses simultaneously and a field gradient pulse techniques we acquired
straightforwardly all pseudo-pure states for two qubits in a single
experiment much efficiently. The signal intensity with the pseudo-pure state
prepared in this way is the same as that of temporal averaging. Our method
is suitable for the system with arbitrary numbers of qubits. As an example
of application, a highly structured search algorithm----Hogg's algorithm was
also performed on the pseudo-pure state $\mid 00\rangle$ prepared by our
method.
\end{abstract}


\vskip 1cm

\narrowtext

\section{INTRODUCTION}

Quantum computers [1-4] use the principle of quantum mechanics and have been
proved to have more advantages over classical computers. Quantum computation
can be divided into three stages consisting of the preparation of initial
states, computation, and the readout of the final states. The proper
preparation of a fiducial pure state as inputs is an important part in the
process. It is not only the starting point of the quantum computation, but
also the foundation of quantum error correction [5-8]. In liquid-state NMR
ensemble quantum computers [9], pseudo-pure states, the highly mixed states,
provide a faithful representation for the transformations of pure states.
Hitherto, several methods [9-16] have been proposed to prepare the
pseudo-pure states including spatial averaging [9,10], temporal averaging
[11], logical labeling [12,13,16], spatially encoding [14] and the cat-state
benchmark [15]. In this paper, we report the experimental implementation of
a new spatial averaging method for pseudo-pure state preparation with
line-selective pulses and a field gradient pulse. We excite specific
single-quantum transitions using selective pulses with appropriate rotating
angles to equalize the populations of the energy levels expect an
undisturbed one. Then a magnetic field gradient pulse along the {\it z-axis}
is exploited to annihilate all transverse magnetizations caused by
line-selective pulses. Possibilities of extending this preparation scheme to
more than two-qubit system are discussed. Advantages and disadvantages of
the scheme are analyzed. A practical example using the pseudo-pure state
prepared with this method for a quantum algorithm is also presented.

\section{NEW METHOD OF PREPARATION OF PSEUDO-PURE STATES}

Consider a system with two spin-1/2 nuclei (AX) with energy levels labeled
as Fig.1. In order to prepare a pseudo-pure state $\mid 00 \rangle$, two
single-quantum transitions irrelevant to the state, that is, the transitions 
$\mid 10 \rangle \leftrightarrow \mid 11\rangle $ for spin X with the pulse
angle $\beta_1$ and $\mid 11\rangle \leftrightarrow \mid 01\rangle$ for spin
A with $\beta_2$ are excited simultaneously. The excitation yields the
operator U [18],

\begin{equation}
U=exp(-i(\beta _1I_x^{(3,4)}+\beta _2^{(4,2)}I))
\end{equation}
where $I_x^{(3,4)}=E_{-}^1I_x^2$, $I_x^{(4,2)}=I_x^1E_{-}^2$ and $%
[I_x^{(3,4)},I_x^{(4,2)}]\neq 0$. Here, $I_\eta ^i(\eta =x,y,z)$ are Pauli
matrices of the {\it i-th} spin, and $E_{\pm }^i=\frac 12(1_2\pm 2I_z^i)$.
The superscripts 1 and 2 stand for the spins A and X respectively. $I_\eta
^{(m,n)}$ are {\it Cartesian} single-transition operators, with {\it m,n}
indicating the energy levels. Simultaneous excitation of two connected
transitions is not straightforward to analyze, but modern computers supply
software for manipulating this operation. In NMR, any pulse excitation
corresponds to a unitary transformation, which can be written into the
matrix form. We choose appropriate values of $\beta _1$ and $\beta _2$
through computer simulation to equalize the populations of the levels 2, 3
and 4. After applying a field gradient pulse {\it Gz} along the {\it z-axis}
to annihilate all transverse magnetizations, the pseudo-pure state $%
|00\rangle $ is finally obtained. The pulse sequence of operations is

\begin{equation}
(\beta _1)_x(\beta _2)_x\rightarrow Gz
\end{equation}
To make it more concrete, we shall show how to find out appropriate $\beta
_1 $ and $\beta _2$ for some specific quantum systems. Firstly, let us
consider a homonuclear two-spin system, in which the gyromagnetic ratio $%
\gamma _i$ of two spins is the same, namely $\gamma _1=\gamma _2=\gamma $.
The deviation density matrix of the system in thermal equilibrium is

\begin{equation}
diag[eq]=[2,0,0,-2]
\end{equation}
where ''$diag$'' represents the diagonal elements of the matrix and the
common factor $\frac{\hbar B\gamma }{2kT}$ is omitted. After the application
of the operator $U$ to Eq.(3), the density matrix will evolve to

\begin{equation}
\rho _t=U\rho _{eq}U^{+}
\end{equation}
In order to get the pseudo-pure state, the following equations 
\begin{equation}
\begin{array}{c}
\rho _t(2,2)=\rho _t(3,3) \\ 
\rho _t(2,2)=\rho _t(4,4)
\end{array}
\end{equation}
should be satisfied. Here, $\rho _t(n,n)$ indicates the -th diagonal element
of the density matrix $\rho _t$. Solving Eq. (5), we get a set of
appropriate solutions $(\beta _1,\beta _2)$ in the range of $(0,360^{\circ
}) $, $(\beta _1,\beta _2)\approx (77.40^{\circ },77.40^{\circ })$.
Therefore, applying the pulse sequence of Eq. (2) to Eq. (3) leads to

\begin{equation}
diag[\rho _{eff}]=[2,-0.6667,-0.6667,-0.6667]=-0.6667diag[E]+2.6667[1,0,0,0]
\end{equation}
where E is a $4\times 4$ unit matrix. It can be seen from Eq. (6) that,
apart from a constant, is actually the pure state $|00\rangle $. For a
heteronuclear two-spin system (i.e. $\gamma _1\neq \gamma _2$) with $^{13}$C
and $^1$H, the gyromagnetic ratio $\gamma _1$ is 1.4048 for $^{13}$C and $%
\gamma _2$ is 5.5857 for $^1$H. The deviation density matrix in thermal
equilibrium is

\begin{equation}
diag[\rho _{eq}]=[6.9905,-4.1809,4.1809,-6.9905]
\end{equation}
where the common factor $\frac{\hbar B}{2kT}$ is also omitted. Applying the
same procedures as the homonuclear system to Eq. (7), we get a set of
solutions $(\beta _1,\beta _2)\approx \left( 127.13^{\circ },186.01^{\circ
}\right) $ . The application of the pulse sequence of Eq. (2) yields finally

\begin{equation}
diag[\rho
_{eff}]=[6.9905,-2.3303,-2.3303,-2.3299]=-2.3303diag[E]+9.3208[1,0,0,0]
\end{equation}
Three other pseudo-pure states$|01\rangle $,$|10\rangle $ and $|11\rangle $
for two-qubit system can be prepared with the similar procedures. The
selective transitions and calculated pulse angles for preparing those states
are listed in Table 1 along with those for $|00\rangle $ state.

Table 1 Parameters for pseudo-pure state preparation with selective pulses %
\tabcolsep 14pt 
\begin{tabular}{|c|c|c|}
\hline
pseudo-pure states & Selective transitions & Pulses angles \\ \cline{2-3}
prepared & Spin A $(^{13}C)$ \hspace{0.4cm} \vline \hspace{0.4cm} Spin X $%
(^1H)$ & $\beta_1$ \hspace{0.4cm} \vline \hspace{0.4cm} $\beta_2$ \\ \hline
$\mid 00 \rangle$ & $\mid 01\rangle \leftrightarrow \mid 11\rangle$ \hspace{%
0.4cm} \vline \hspace{0.4cm} $\mid 11\rangle \leftrightarrow \mid 10\rangle$
& 127.13$^{o}$ \hspace{0.4cm} \vline \hspace{0.4cm} 186.01$^{o}$ \\ \hline
$\mid 01 \rangle$ & $\mid 00\rangle \leftrightarrow \mid 10\rangle$ \hspace{%
0.4cm} \vline \hspace{0.4cm} $\mid 10\rangle \leftrightarrow \mid 11\rangle$
& 127.13$^{o}$ \hspace{0.4cm} \vline \hspace{0.4cm} 186.01$^{o}$ \\ \hline
$\mid 10 \rangle$ & $\mid 11\rangle \leftrightarrow \mid 01\rangle$ \hspace{%
0.4cm} \vline \hspace{0.4cm} $\mid 01\rangle \leftrightarrow \mid 00\rangle$
& 127.13$^{o}$ \hspace{0.4cm} \vline \hspace{0.4cm} 186.01$^{o}$ \\ \hline
$\mid 11 \rangle$ & $\mid 10\rangle \leftrightarrow \mid 00\rangle$ \hspace{%
0.4cm} \vline \hspace{0.4cm} $\mid 00\rangle \leftrightarrow \mid 01\rangle$
& 127.13$^{o}$ \hspace{0.4cm} \vline \hspace{0.4cm} 186.01$^{o}$ \\ \hline
\end{tabular}
\vspace{3cm}

The method outlined above can be, in principle, applied to systems with more
than two qubits. For a three-qubit system with energy levels labeled in
Fig.5. a route to cascade the other energy levels except the $|000\rangle $
one, $\mid 010\rangle \stackrel{1}{\longleftrightarrow} \mid 110\rangle 
\stackrel{2}{\longleftrightarrow} \mid 100\rangle \stackrel{3}{%
\longleftrightarrow} \mid 101\rangle \stackrel{4}{\longleftrightarrow} \mid
111\rangle \stackrel{5}{\longleftrightarrow} \mid 011\rangle \stackrel{6}{%
\longleftrightarrow} \mid 001\rangle$ can be utilized to apply
simultaneously selective pulses. The pulse angles $(\beta _1,\beta _2,\beta
_3,\beta _4,\beta _5,\beta _6)$ exciting the corresponding transitions
labeled by the numbers on the symbol are determined by equalizing the
populations of all the levels involved. For a homonuclear system, they were
calculated to be $(\beta _1,\beta _2,\beta _3,\beta _4,\beta _5,\beta
_6)\approx (182.02^{\circ },179.04^{\circ },229.38^{\circ },193.46^{\circ
},200.28^{\circ },105.75^{\circ })$. For a specific heteronuclear system
consisting of the $^{13}$C, $^{13}$C and $^1$H, calculation resulted to $%
(\beta _1,\beta _2,\beta _3,\beta _4,\beta _5,\beta _6)\approx
(201.89^{\circ },258.83^{\circ },313.40^{\circ },346.31^{\circ
},295.37^{\circ },234.18^{\circ })$. A pseudo-pure state for a three-qubit
system is thus achieved after a field gradient pulse is applied. If one
could execute specific pectinate pulses and control their excitation
intensity as required by developing NMR experimental technique and the
instrumental function, the method can be extended to many-qubit systems.

\section{EXPERIMENTAL NMR RESULTS}

The scheme stated above was implemented by liquid-state NMR spectroscopy
with carbon-13 labeled chloroform $^{13}$CHCl$_3$ (Cambridge Isotope
Laboratories, Inc.). We chose H and C as the two-spin system in the
experiments. Spectra were recorded on a BrukerARX500 spectrometer with a
probe tuned at 125.77MHz for $^{13}$C (denoted by A), and at 500.13MHz for
the $^1$H(denoted by X). The spin-spin coupling constant J between $^{13}$C
and $^1$H is 214.95Hz. The relaxation times were measured to be T1=4.8sec
and T2=0.2sec for the proton, and T1=17.2sec and T2=0.35sec for carbon
nuclei.

The $|00\rangle $ pseudo-pure state was firstly prepared in the experiment.
Selective excitations were executed using low-power, long-duration pulses of
a {\it guass1k} shape. The length of these pulses was tailored to achieve
sufficient selectivity in the frequency domain without disturbing the
nearest line, depending on the magnitude of the J coupling constant. We took
all the length of the selective pulses to be 32ms in order to meet the
requirement of simultaneity. The excitation powers for the channels $^{13}$C
and $^1$H were set to be 56.7dB and 63.1dB so that the rotating angles of
and were obtained respectively. The magnetic field gradient pulse was
accomplished with the {\it sine-0} shape along the z axis. The spectra then
recorded (shown in Fig.2) confirm that a pseudo-pure state $|00\rangle
\langle 00|$ has been prepared. Fig.3 (a) and (b) show respectively the real
and imaginary components of the deviation matrix of for the $|00\rangle $
state by quantum state tomography. The pseudo-pure state we obtained was

\vspace{0.8cm} $\rho =\left[ 
\begin{array}{cccc}
1 & -0.0096-0.0090i & 0.0017-0.0026i & -0.0005+0.0006i \\ 
-0.0096+0.0090i & 0.0230 & -0.0020+0.0014i & 0.0033+0i \\ 
0.0017+0.0026i & -0.0020-0.0014i & 0.0161 & 0.0095+0.0090i \\ 
-0.0005-0.0006i & 0.0033-0i & 0.0095-0.0090i & 0
\end{array}
\right] $

The maximal relative error of the experimental values of the density matrix
elements was 
\mbox{$<$}%
3\%.Applying the corresponding transition-selective pulses given in Table 1
we got all the other pseudo-pure states $|01\rangle $, $|10\rangle $ and $%
|11\rangle $ for the two-spin system. The experimental results are shown in
Fig.3.(c)-(h) with the maximal experimental error 
\mbox{$<$}%
5\%.

\section{AN APPLICATION TO HOGG ALGORITHM}

We experimentally implemented the Hogg's algorithm [19,20] with the
pseudo-pure state $|00\rangle $ we prepared above. For a two-qubit system,
the solution to the 1-SAT problem with two clauses was to be sought after.
The logic formula is$V_{1}\bigwedge V_{2}$, ($V_{1},V_{2}$ are the logic
variables). The corresponding solution is theoretically found to be $|11$ $%
\rangle $ . Applying the sequence in [20] to the pseudo-pure state $%
|00\rangle $, we got the results shown in Fig.4 with the experimental errors 
\mbox{$<$}%
5\%.

\section{DISCUSSIONS}

In comparison with previous methods for preparing pseudo-pure states, the
present method using line-selective excitations seems to be simpler and
experimentally more efficient,. With temporal averaging [11], the
preparation of a two-spin pseudo-pure state needs to carry out three
different experiments. For spatial averaging proposed in [4], the
magnetization are greatly lost due to the repeated magnetic field gradient
pulses introduced in the course of the preparation so that the signal
intensity is reduced. Moreover, as logical labeling[12,13,16] requires some
qubits as ancillary bits, which don't participate in computation, much
memory room is wasted. Furthermore its Signal-to-Noise ratio(SNR) is not
ideal due to SNR$\propto nvN\alpha /2^N$ [12], with {\it n} being the
molecular density of the sample, {\it v} the volume, and {\it N} the spin
numbers of the sample molecule. In contrast, using the present method, one
can apply fewer pulses to prepare a pseudo-pure state in a single experiment
with the signal intensity of $\frac 23(\gamma _1+\gamma _2)$, the same as
that of temporal averaging. Moreover, all pseudo-pure states for a specific
spin system can be obtained with a judicious choice of line-selective
pulses. Besides the errors due to the inhomogeneity of RF fields and static
magnetic fields as well as imperfections of the pulse-length calibration,
another main error in experiments is introduced from the imperfections of
the line-selective pulses. Selective excitation of a {\it guass1k} shape
leads to a large phase gradient of lines in excitation band. Theoretically,
the phase gradient can be corrected in some way, but much base-line
distortion will practically be caused. However, from the experimental
results, we can find that the errors caused by the present method is smaller
than that by previous ones. A difficulty of the present method is related to
the sample choice. As line-selective pulses require all lines to be
well-resolved, the coupling constants J should increase rapidly with the
increase of the qubit number, but are still much smaller than chemical
shift. It may be very hard to find out or synthesize such a sample for
many-qubit system.

\begin{center}
{ACKNOWLEDGEMENTS}
\end{center}

We thank Xiaodong Yang, Tao Zhang, Fei Du, Hanzheng Yuan, Yonghong Zhang and
Xu Zhang for help in the course of experiments.

\newpage

\begin{center}
{\bf Captions of the figures}
\end{center}

Figure 1~~The energy levels of a two-spin system (each with spin-$\frac{1}{2}
$ ) and the scheme of preparing the pseudo-pure state $|00\rangle $.

Figure 2~~Spectra of the prepared pseudo-pure state $|00\rangle $. The left
indicates the spectra of $^{13}$C and the right, $^1$H. The reading-out
pulses (a) , (b) and (c) were applied respectively. The abscissa indicates
the frequency, and the ordinate denotes the intensity of the spectra (in
arbitrary unit).

Figure 3~~Experimentally measured deviation density matrices for the
pseudo-pure states, denoting the $|00\rangle $,$|01\rangle $,$|10\rangle $%
and $|11\rangle $ states respectively from top to bottom. The left and right
column show the real and imaginary components of those $|00\rangle $,$%
|01\rangle $,$|10\rangle $and $|11\rangle $ states respectively.

Figure 4~~Experimental results of the final density matrix after
implementation of the Hogg's algorithm of logical formula V$_1\bigwedge $V$_2
$ on the pseudo-pure state $|00\rangle $. (a) the real components. (b) the
imaginary components.

Figure 5~~The energy levels of a three-spin system (each with spin-$\frac{1}{%
2}$ ) and the scheme of preparing the pseudo-pure state $|000\rangle $.

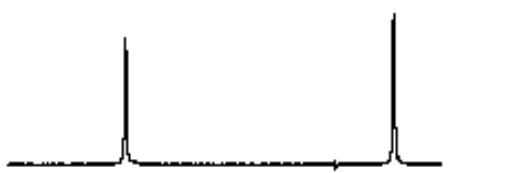
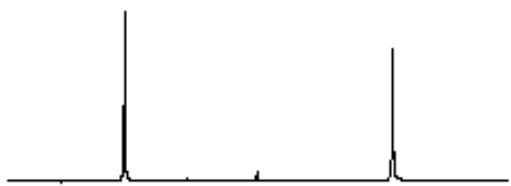
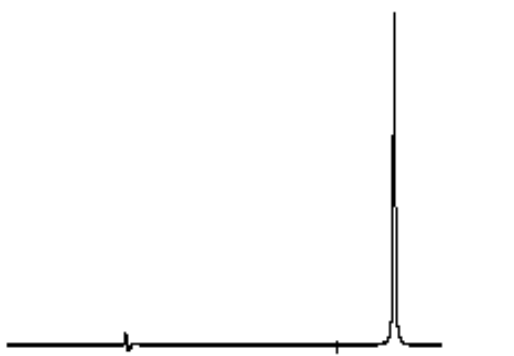
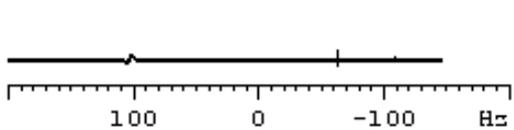
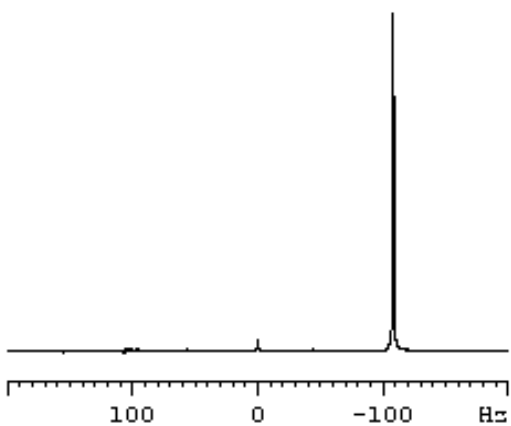
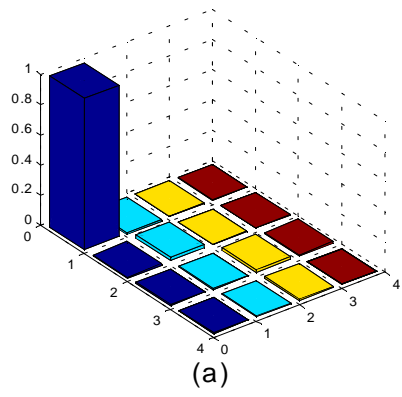
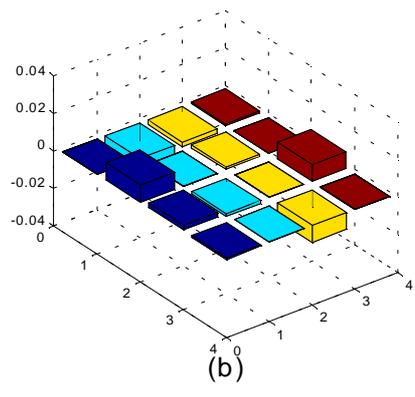
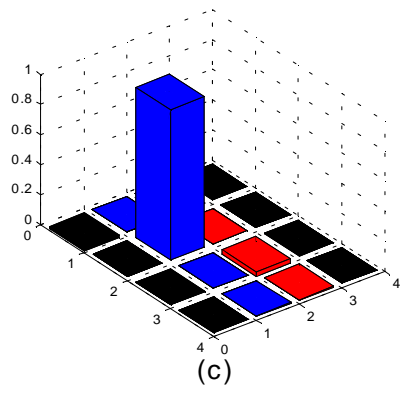
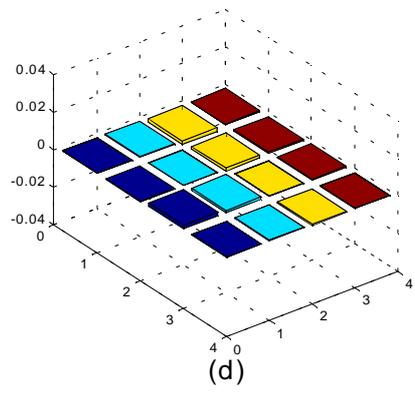
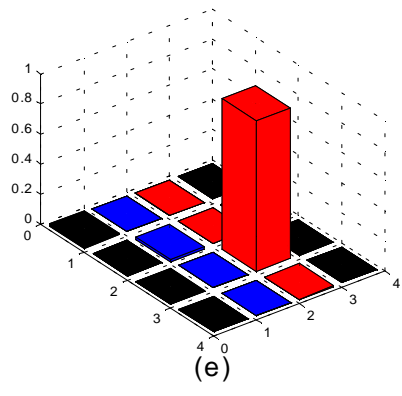
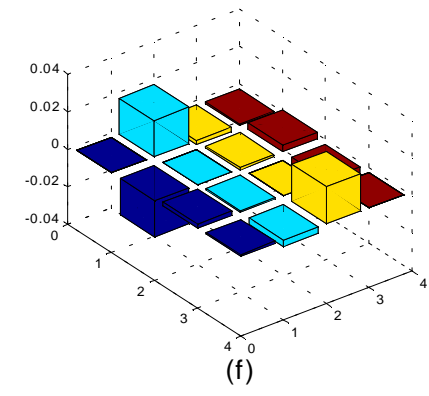
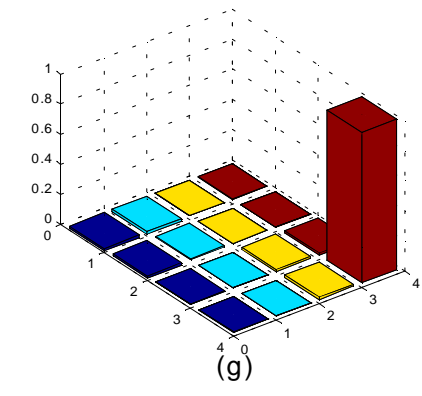
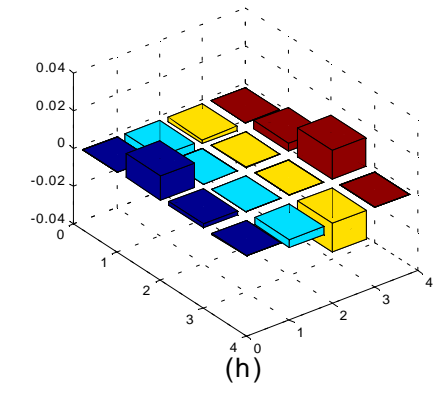
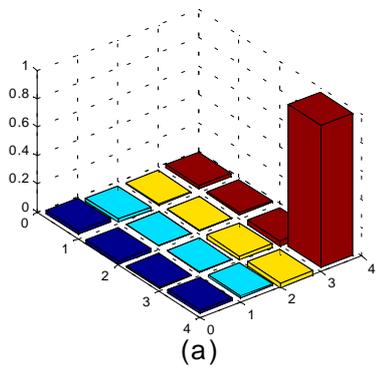
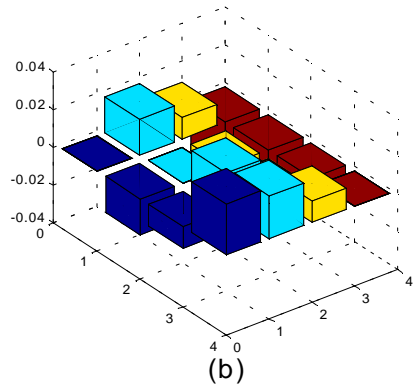
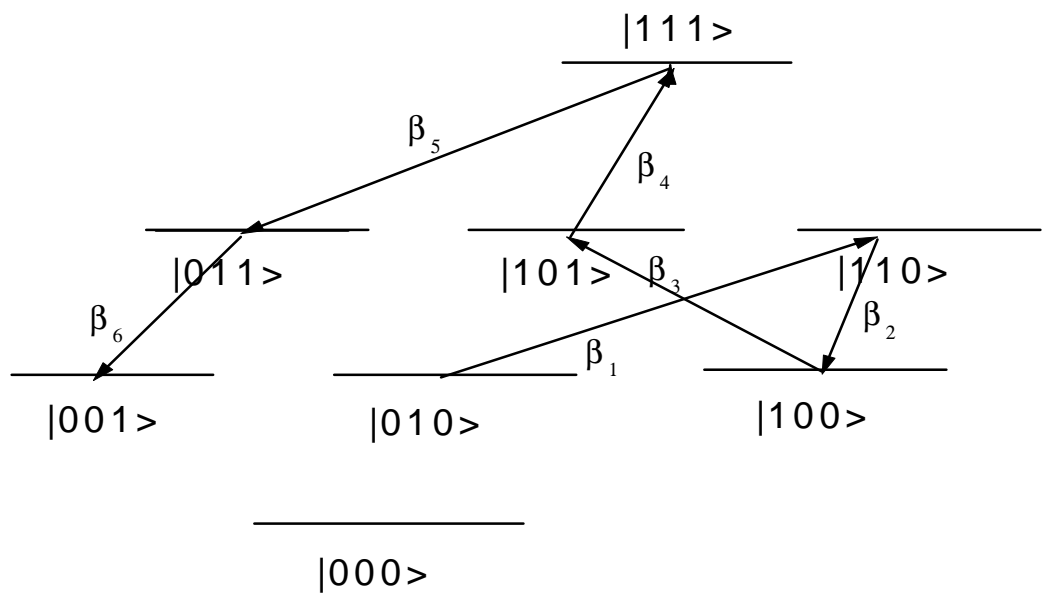
\end{document}